\documentclass[11pt,a4paper,english,twoside]{article}

\usepackage{a4wide}
\usepackage{amssymb, amsmath}
\usepackage{graphicx}
\usepackage[all]{xy}
\usepackage{hyperref}
\hypersetup{linktocpage}
\usepackage{enumerate}
\usepackage{xcolor}
\usepackage{dsfont}
\usepackage{empheq}
\usepackage{cite}
\usepackage{float}

\usepackage[utf8]{inputenc}
\usepackage{soul}
\usepackage{subcaption}

\newcommand{\beq}{\begin{equation}}
\newcommand{\eeq}{\end{equation}}
\def\bea#1\eea{\begin{align}#1\end{align}}
\def\beal#1\eeal{\begin{subequations}\begin{align}#1\end{align}\end{subequations}}
\newcommand{\nn}{\nonumber}
\newcommand{\w}{\wedge}

\newcommand{\eq}[1]{\begin{equation}#1\end{equation}}
\newcommand{\spl}[1]{\begin{split}#1\end{split}}

\newcommand{\swed}{{\scriptscriptstyle \wedge}}

\def\d {{\rm d}}

\begin{document}
\numberwithin{equation}{section}

\begin{titlepage}

\begin{flushright}    
  {\small 
  LMU-ASC 15/20 \\
  MPP-2020-50
  }
\end{flushright}

\begin{center}

\phantom{DRAFT}

\vspace{1.0cm}

{\LARGE \bf{AdS$_2$ Type-IIA Solutions and Scale Separation}}\\

\vspace{2 cm} {\Large Dieter L\"ust$^{1,2}$ and Dimitrios Tsimpis$^{3}$}\\
 \vspace{0.9 cm} {\small\slshape $^1$ Arnold-Sommerfeld-Center for Theoretical Physics\\
Ludwig-Maximilians-Universit\"at, 80333 M\"unchen, Germany}\\
 \vspace{0.2 cm} {\small\slshape $^2$ Max-Planck-Institut f\"ur Physik (Werner-Heisenberg-Institut)\\
 F\"ohringer Ring 6, 80805, M\"unchen, Germany}\\
 \vspace{0.2 cm} {\small\slshape $^3$ 
 Institut de Physique des Deux Infinis de Lyon\\
Universit\'{e} de Lyon, UCBL, UMR 5822, CNRS/IN2P3\\
4 rue Enrico Fermi, 69622 Villeurbanne Cedex, France}\\
\vspace{0.5cm} {\upshape\ttfamily dieter.luest@lmu.de; tsimpis@ipnl.in2p3.fr}\\

\vspace{3cm}

{\bf Abstract}
\vspace{0.1cm}
\end{center}

\begin{quotation}
\noindent 
In this note we examine certain classes of solutions of IIA theory without sources, of the form 
AdS$_2\times {\cal M}^{(1)}\times \dots \times {\cal M}^{(n)}$, where  ${\cal M}^{(i)}$ are Riemannian  spaces. 
We show that large hierarchies of curvatures can be obtained between the different factors, 
however the absolute value of the scalar curvature of AdS$_2$ must be of the same order or larger  than the absolute values of the scalar curvatures of  all the other factors.

\end{quotation}

\end{titlepage}

\newpage



\tableofcontents

\section{Introduction}\label{sec:intro}

{The question of scale separation of AdS vacua in string theory \cite{tss,ss2,ss3} is the subject to a lot of recent discussion. 
In many of the previous works, backgrounds of the form AdS$_d\times {\cal M}^{(p)}$ were considered, and special emphasis was given to the case with $d=4$ and $p=6$.
Scale separation is possible whenever the radius $L_d$ of  AdS$_d$ is parametrically larger than the inverse Kaluza-Klein (KK) mass scale $L_p\simeq \tfrac{1}{m_{\text{KK}}}$ of ${\cal M}^{(p)}$, i.e. $L_d\gg  L_p$. In the case where scale separation is possible, the theory possesses a limit in which the solution can be regarded as $d$-dimensional AdS$_d$ space. On the other hand, if scale separation is not possible, 
i.e. if $L_d\simeq L_p$, the solutions are not really $d$-dimensional, and the gravitational background has to be considered as $(d+p)$-dimensional. 
Whether scale separation in supergravity backgrounds of the above form is possible or not has also profound consequences for the holographically dual CFT in $(d-1)$ dimensions.}

{We do not want to review all arguments which were given in favor or against scale separation for AdS$_d\times {\cal M}^{(p)}$ background spaces.
Some general arguments against scale separation were given in \cite{ss2}. 
On the other hand, one of the early papers addressing this issue is the work of DGKT \cite{DeWolfe:2005uu}, where it was argued that in the presence of orientifold planes scale separation is
possible. This discussion was recently refined and extended in \cite{Blumenhagen:2019vgj,ss4,Apruzzi:2019ecr,Junghans:2020acz,Buratti:2020kda,Marchesano:2020qvg}.
The question of scale separation was also recently addressed in the general context of the quantum gravity swampland discussion \cite{Vafa:2005ui}, namely as the AdS Distance 
Conjecture (ADC) \cite{Lust:2019zwm}. This conjecture 
states that the limit of small AdS cosmological constant, $\Lambda\simeq \frac{1}{L_d^{2}}\rightarrow0$,  is at infinite distance in the space of AdS metrics, and that it is related to an infinite tower of states 
with typical masses that scale as,
\begin{equation}
{\rm ADC}:\quad m\sim\Lambda^{\alpha}\, ,
\end{equation}
with $\alpha={\cal O}(1)$.  The strong version of the ADC proposes that for supersymmetric backgrounds  $\alpha=\tfrac12$,
and that in this case scale separation is not possible, since $L_d\sim \tfrac{1}{m}$.}

\vskip0.3cm
In this paper we will consider several AdS$_2$ solutions in string theory, where the total space is of the form
AdS$_2\times {\cal M}^{(1)}\times \dots \times {\cal M}^{(n)}$. 
Assuming that the scale, or radius, is related to the scalar curvature, or cosmological constant, via $\Lambda\simeq\frac{1}{L^2}$, we will see that scale separation for these backgrounds will never be possible
in the sense that for all the considered cases the radius $L_2$ of AdS$_2$ can never be much larger than  {at least one of the radii }of    the other factors.
E.g. if ${\cal M}^{(1)}$ is a two-sphere $S^2$ of radius $L_2'$, then  $L_2\leq L_2'$. However there are cases where the rest of the radii, e.g. the radius of a Ricci-flat space $M_{(6)}$,
can be much smaller than $L_2$, $L_2'$. In particular,  $L_2=L_2'\gg L_6$ is possible. This means that there can be scale separation between AdS$_2\times S^2$ and $M_{(6)}$, even within a regime of weak coupling and curvature where  the supergravity approximation is valid. The reason this is possible is that in the Ricci-flat case the radius is no longer related to the inverse of the scalar curvature (which vanishes). Instead the radius becomes a free parameter of the solution, only constrained by flux quantization.

{The case of AdS$_2\times S^2$ is of special interest, since it corresponds to the near horizon geometry of four-dimensional extremal, supersymmetric black holes.
The radii $L_2= L_2'$ are directly related to the entropy ${\cal S}$ of the corresponding black hole solutions: 
\begin{equation}
{\cal S}\sim L_2^2\, .
\end{equation}
As it was recently discussed in \cite{Bonnefoy:2019nzv}, the limit of large entropy, ${\cal S}\rightarrow\infty$, is at infinite distance in the space of 4D black hole metrics. Therefore, similarly to the ADC, 
 a black hole entropy conjecture (BHEC) was put forward  in \cite{Bonnefoy:2019nzv}, stating that the large entropy limit of black holes is also accompanied by a tower of light modes. However these modes cannot be given in terms
of the internal KK modes of $M_{(6)}$. This was already 
seen in \cite{Bonnefoy:2019nzv}  from the so-called attractor equations, since, as a function of the electric and magnetic black hole charges, ${\cal S}$ can be made large, while keeping the internal scale $L_6$ small.
Here we will confirm this result by investigating the supergravity solutions of the corresponding intersecting D-branes, and reading off  from the supergravity solutions the corresponding length scales. 
However, as we will discuss, there are other classes of AdS$_2$ supergravity solutions, where scale separation is only possible in the other direction: namely there are solutions where
 the AdS$_2$ space 
 and some of the internal factors are more highly curved than the rest of the internal factors. We call this the ``{\sl wrong}'' kind of scale separation.
}

\vskip0.2cm

{The paper is organized as follows. In the next section we will briefly review the background spaces of supergravity $p$-branes in ten spacetime dimensions.
In \S\ref{sec:rn} we will then discuss the construction of supergravity solutions of intersecting D-branes, which lead to supersymmetric 4D black holes with AdS$_2\times S^2$  near-horizon geometry. 
We will see that scale separation $L_2=L_2'\gg L_6$ is possible within the validity regime of the  supergravity approximation. 
We also compare these results with those of \cite{Bonnefoy:2019nzv}. 
In \S\ref{sec:gen}, we discuss various generalizations of spaces $M_6$ and show that scale separation works in a different way than before. 
The  case of $M_6$  Ricci-flat,  discussed in \S\ref{sec:rf},   includes the brane set up of \S\ref{sec:nh} as a special case.
In \S\ref{sec:45} the case of backgrounds of the form $M_{10}=M^{(1)}_{2}\times\dots\times M^{(5)}_{2}$ is analyzed. In \S\ref{sec:fq}
we discuss solutions of the  form AdS$_2\times $S$^2\times$S$^2\times$S$^2\times$T$^2$, 
AdS$_2\times $S$^2\times$S$^2\times$T$^4$, or AdS$_2\times $S$^2\times$S$^2\times$K3, obeying flux quantization within the validity regime of the  supergravity approximation. 
We conclude with a discussion  in \S\ref{sec:discussion}. 
}

\section{$p$-branes in $D=10$}\label{sec:brane10}

For a review of brane solutions see e.g.~\cite{Duff:1993ye,Youm:1997hw,Johnson:2000ch}. 
In ten dimensions,  in the string frame, $p$-branes  are  solutions of the action $S=S_{{\rm bulk}} + S_{{\rm sources}}$ where,  
\beq
S_{{\rm bulk}} = \frac{1}{2\kappa_{10}^2} \int \d^{10} x \, \sqrt{|g_{10}|}\ \left( e^{-2\phi} ({\cal R} +4 |\d \phi|^2 ) - \frac{1}{2} |F_{p+2}|^2 \right)\ ,
\eeq
$g_{10}$ is the determinant of the  metric $g_{MN}$, $M,N= 0 \dots 9$,   $F_{p+2} = \d C_{p+1}$ 
is the abelian $(p+2)$-form field strength, and $\phi$ is the dilaton. The square of a $q$-form  $A_q$ is 
defined by
$|A_q|^2 = A_{q\, M_1 \dots M_q}\, g^{M_1N_1} \! \dots g^{M_q N_q} A_{q\, N_1 \dots N_q} / q!$. 
Moreover,
\beq
S_{{\rm sources}} = - T_p \int_{\Sigma_{p+1}} \d^{p+1}\xi \ e^{-\phi} \sqrt{|\imath^*[g_D]|} + \mu_p \int_{\Sigma_{p+1}} \imath^*[C_{p+1}] \ ,
\eeq
where $\Sigma_{p+1}$ is the  world-volume of the $p$-brane with coordinates $\xi^i$, $i=0 \dots p$, and $\imath^*[ \cdot]$ the pull-back to $\Sigma_{p+1}$. 
The gravitational constant and the tension are given  by,
\beq
\label{tension}
2\kappa_{10}^2=(2\pi)^7 (\alpha^\prime)^4\ , \quad T_p^2=\frac{\pi}{\kappa_{10}^2} (4\pi^2 \alpha^\prime)^{3-p} \ ,
\eeq
where   $\alpha^\prime=l_s^2$, with $l_s$ the string length. For BPS sources as here, one has  $\mu_p=T_p$. 
The $p$-brane solutions in  the  string frame are then given by,
\eq{\spl{\label{solgenstring}
\d s^2&= H^{-\frac{1}{2}} \eta_{ij}\d x^i\d x^j+ H^{\frac{1}{2}} \delta_{mn}\d y^m\d y^n\\
e^\phi&=e^{\phi_0}\, H^{-\frac{(p-3)}{4}}~;~~~C_{{p+1}}=(H^{-1} -1)\, e^{-\phi_0}\, \text{vol}_{p+1}
~,}}
with $x^{i=0 \dots p}$ the coordinates along the brane,  $\text{vol}_{p+1} = \d x^0 \w \dots \w \d x^p$, and $y^{m=p+1 \dots 9}$ 
the coordinates of the space transverse to the brane. 
$H(\vec{y})$ is a  harmonic function with localized source in the (unwarped) $\mathbb{R}^{9-p}$ space transverse to the $p$-brane,
\eq{\label{laplpbrane}
\delta^{mn}\partial_m\partial_n H(\vec{y})=  {Q}\, \delta(\vec{y}-\vec{y}_0)
~,}
where the brane is located  at $\vec{y}_0$ in the transverse space.
For $D_p$-branes we have,
\beq
\label{52}
Q_{D_p}= - 2 \kappa^2_{10} T_p g_s = -(2\pi l_s)^{7-p} g_s  \ .
\eeq
The constant $e^{\phi_0}=g_s$ can be used to define the string coupling 
 as the value of the dilaton at asymptotic infinity, where the harmonic functions tend to unity. 
 However, once the near-horizon limit is taken (see section \ref{sec:nh}), the asymptotic region is no longer accessible. 
More generally  one should think of $g_s$ as a free parameter of the solution, related to the string coupling via \eqref{solgenstring}.

\section{The brane configuration}\label{sec:rn}

One can form superpositions of  brane solutions according to the 
{\it harmonic superposition rule}~\cite{hs1,hs2}.  
Consider the
following system of intersecting D4/D0-branes:
\bigskip
\begin{center}
  \begin{tabular}{|c||c|c|c|c|c|c|c|c|c|c|}
    \hline
    & $t$ & $x^1$  & $x^2$  & $x^3$ & $y^1$  & $y^2$  & $y^3$  & $y^4$  & $y^5$  & $y^6$ \\
    \hline
    \hline
    $\mathrm{D}4_{1}$ & $\bigotimes$ &  &   &  &$\bigotimes$  &  $\bigotimes$ &  $\bigotimes$  & $\bigotimes$ 
                &   &  \\

    \hline
    $\mathrm{D}4_{2}$ & $\bigotimes$ &    &  &  &  &  & $\bigotimes$  &
               $\bigotimes$  &  $\bigotimes$  & $\bigotimes$   \\

    \hline
    $\mathrm{D}4_{3}$ & $\bigotimes$ &  &  &   & $\bigotimes$   & $\bigotimes$  &   &
                 & $\bigotimes$  & $\bigotimes$ \\
 \hline
    $\mathrm{D}0_{\phantom{1}}$ & $\bigotimes$ &    &  &   &    &   &   &   
                 &    &   \\
\hline
   \end{tabular}
 \end{center}
\bigskip
where $y^{m=1,\dots,6}$ are assumed to parameterize a $T^6$, and $x^{i=1,2,3}$ are 
coordinates of $\mathbb{R}^3$. 
We use the notation  D4$_{\alpha}$, $\alpha=1,2,3$ to distinguish the three different types of D4-branes in the configuration of the table above. 
We shall assume there are  $n_0$  D0-branes located at $\vec{x}_{0}\in\mathbb{R}^3$, and  $n_{\alpha}$  D4$_{\alpha}$-branes located at 
$\vec{x}_{\alpha}$, $\alpha=1,2,3$.

The explicit form of the metric  reads, 
\begin{align}
\label{metric}
\mathrm{d}s_{10}^2=& -\Big(\prod_{\alpha=0}^3
H_{\alpha} \Big)^{-\frac{1}{2}}~
 \d t^2
 +\Big(\prod_{\alpha=0}^3
H_{\alpha} \Big)^{\frac{1}{2}}~
\sum_{i=1}^3(\d x^i)^2
+\sqrt{\frac{H_{0} H_{2}}{ H_{1}H_3} }\left(
 (\mathrm{d}y^1)^2
+ (\mathrm{d}y^2)^2
\right)
 \nn\\
 &+\sqrt{\frac{H_{0}H_{3}}{H_{1}H_2}}
\left(
 (\mathrm{d}y^3)^2
+ (\mathrm{d}y^4)^2 \right)
+\sqrt{\frac{H_{0}H_{1}}{H_{2}H_3}}
\left(
 (\mathrm{d}y^5)^2
+ (\mathrm{d}y^6)^2\right)
 ~,
\end{align}
where $H_{\alpha}$, $\alpha=1,2,3$, are the harmonic functions of the D4$_{\alpha}$-branes, and $H_0$ 
is the  harmonic function of the   D0-branes. 
We  have,
\eq{
H_{\alpha}=1+\frac{c_{\alpha}}{|\vec{x}-\vec{x}_{\alpha}|}~;~~~
c_{\alpha}=\frac{N_{\alpha}  g_s} {4\pi}  (2\pi l_s)^{7-p} 
~,}
for $\alpha=0,\dots, 3$, and we took into account that \eqref{laplpbrane} implies $c=-\frac{Q}{4\pi}$, for the case where the transverse space is $\mathbb{R}^3$. 
The $N_\alpha$ are proportional to the number of D-branes $n_\alpha$. The precise relation will be derived below using flux quantization.\footnote{In the case of a single set of parallel D-branes, we would simply have $N_\alpha=n_\alpha$. However the brane solution \eqref{metric} was obtained using harmonic superposition, which results in smearing the D-branes along the directions of the $T^6$. As a consequence,    $N_\alpha$, $n_\alpha$ are not equal to each other.}  
More explicitly, 
\eq{\label{csts}
c_0= \frac{N_{0}  g_s} {4\pi}  (2\pi l_s)^{7}~;~~~
c_\alpha=\frac{N_{\alpha}  g_s} {4\pi}  (2\pi l_s)^{3}~,~\alpha=1,2, 3
~.}

\subsection{Near-horizon limit}\label{sec:nh}

We shall now assume that all branes are located at the origin: $\vec{x}_{\alpha}=0$, $\alpha=0,\dots,3$. 
Let us define $r:=\sqrt{\vec{x}^2}$. In the {\it near-horizon} limit $r\rightarrow 0$, \eqref{metric} reads,
\begin{align}
\label{metricnh}
\frac{1}{C}~\! \mathrm{d}s_{10}^2=& 
-r^2\d t^2+\frac{\d r^2}{r^2}+\d\Omega^2\nn\\
&+{\frac{1}{ c_{1}c_3} }\left(
 (\mathrm{d}y^1)^2
+ (\mathrm{d}y^2)^2
\right)
 +{\frac{1}{c_1c_2}}
\left(
 (\mathrm{d}y^3)^2
+ (\mathrm{d}y^4)^2 \right)
+{\frac{1}{c_2c_3}}
\left(
 (\mathrm{d}y^5)^2
+ (\mathrm{d}y^6)^2\right)
 ~,
\end{align}
where $\d\Omega^2$ 
is the line element of the unit two-sphere. Moreover 
we  
defined $C:= \big(\prod_{\alpha=0}^3
c_{\alpha} \big)^{\frac{1}{2}}$, and 
rescaled the time coordinate: $t\rightarrow t/C$.

The fluxes read, 
\eq{\spl{\label{flux}
g_sF_2&= 
\frac{C}{c_0}\d r\wedge\d t\\
g_s F_6&=C\d r\wedge\d t \wedge\Big(\frac{1}{c_1}\d y^1\wedge\d y^2 \wedge\d y^3\wedge\d y^4\\
&+\frac{1}{c_2}  \d y^3\wedge\d y^4\wedge\d y^5\wedge\d y^6
+\frac{1}{c_3}  \d y^1\wedge\d y^2 \wedge\d y^5\wedge\d y^6
\Big)\\
g_s\star F_2&=c_0\d\Omega_2\wedge\d y^1\wedge\dots\wedge\d y^6\\
g_s \star F_6&= \d\Omega_2\wedge\Big(
 {c_1}\d y^5\wedge\d y^6  
+ {c_2}  \d y^1\wedge\d y^2 
+ {c_3}  \d y^3\wedge\d y^4  
\Big)
~,
}}
where $\d\Omega_2$ is the volume form of the unit 2-sphere, and we have taken into account that the time coordinate has been rescaled as indicated below \eqref{metricnh}. 
All fluxes, as well as their Hodge-duals, can readily be seen to be everywhere well-defined and closed, $\d F=\d\star F=0$, indicating the absence of sources. 
In other words, the near-horizon limit is a pure gravity background, all branes having dissolved into fluxes in the limit. 

We shall assume that the areas, in units of string length,  of the three 2-tori $\Sigma_i$ spanned by the internal coordinates are given by three moduli $v_i$, 
\eq{\label{moduli}
v_1=\frac{1}{l_s^2}\int_{\Sigma_1}\d y^1\d y^2~;~~~
v_2=\frac{1}{l_s^2}\int_{\Sigma_2}\d y^3\d y^4~;~~~
v_3=\frac{1}{l_s^2}\int_{\Sigma_3}\d y^5\d y^6
~.}
The flux quantization conditions,  
\eq{\spl{\label{fc}
n_0&=\frac{1}{(2\pi l_s)^{7}}\int_{S^2\times T^6} \star F_2 ~;~~~\\
n_1&=\frac{1}{(2\pi l_s)^{3}}\int_{S^2\times \Sigma_3} \star F_6  ~;~~~
n_2=\frac{1}{(2\pi l_s)^{3}}\int_{S^2\times \Sigma_1} \star F_6~;~~~
n_3=\frac{1}{(2\pi l_s)^{3}}\int_{S^2\times \Sigma_2} \star F_6 
~,}}
then relate $N_\alpha$ to the number of D-branes $n_\alpha \in \mathbb{N}$, which have dissolved into flux quanta in the near-horizon geometry,
\eq{\label{nnrel}
N_0=\frac{n_0}{l_s^6 v_1 v_2 v_3} ~;~~~
N_1=\frac{n_1}{l_s^2 v_3}~;~~~
N_2=\frac{n_2}{l_s^2 v_1}~;~~~
N_3=\frac{n_3}{l_s^2 v_2} 
~,
}
where we have substituted \eqref{flux} into \eqref{fc}, taking \eqref{csts}, \eqref{moduli} into account.

The geometry of \eqref{metricnh} is AdS$_2\times$S$^2\times$T$^6$. 
The radii $L_2$, $L_2'$ of AdS$_2$, S$^2$ respectively, can be seen to be equal to each other. 
The 4D part of the  geometry, AdS$_2\times$S$^2$, is thus characterized by a radius $L_4:= L_2= L_2'$. 
The latter and the radius $L_6$ of T$^6$ can be read off of \eqref{metricnh}, 
\eq{\spl{
L_4&=C^{\frac12}
=4\pi^3l_s g_s(n_0n_1n_2n_3)^{\frac14}(v_1v_2v_3)^{-\frac12}
~;~~~\\
L_6&=\frac{C^{\frac12}}{(c_1c_2c_3)^{\frac13}}l_s(v_1v_2v_3)^{\frac16}=
2\pi l_s\Big(
\frac{n_0^3}{n_1n_2n_3}
\Big)^{\frac{1}{12}}
~.}}
Unlike the individual values of the radii $L_4$, $L_6$, their ratio is frame-independent, 
\eq{\label{sr}
\frac{L_4}{L_6}=
 { 2\pi^2 ~\!g_s}
(n_1n_2n_3)^{\frac13}(v_1v_2v_3)^{-\frac12}
~.}
The dilaton is constant, 
\eq{ \label{dilaton}
e^\phi=g_s(2\pi)^3\Big(
\frac{n_0^3}{n_1n_2n_3}
\Big)^{\frac{1}{4}}(v_1v_2v_3)^{-\frac12}
 ~.
}

\subsection{Validity}

The metric,  fluxes and dilaton \eqref{metricnh}, \eqref{flux}, \eqref{dilaton} give an exact bulk supergravity solution without sources. The solution is parameterized by  the parameters 
$g_s$, $n_a$, $v_i$, which can be tuned independently. 

Let us denote by $T_i$ the effective areas (in string frame and in string units) of the three 2-tori in the near-horizon limit, 
\eq{\label{ea}
T_2:=\frac{C}{c_1c_3}  v_1~;~~~
T_3:=\frac{C}{c_1c_2} v_2~;~~~
T_1:=\frac{C}{c_2c_3} v_3
~.}
Taking \eqref{csts}, \eqref{nnrel} into account, this is equivalent to, 
\eq{\label{deft}
T_i=4\pi^2n_i\sqrt{  \frac{n_0}{n_1n_2n_3}  } ~.}
For the validity of the supergravity regime we must require, 
\eq{\label{cond1}
T_i\gg 1~.} 
In addition we must require that 
the radius of curvature of the background is much larger than the 
string scale,
\eq{
\label{cond2}
L_4, L_6\gg l_s ~,}
and that the string coupling obeys 
\eq{
\label{cond3}
e^\phi  <1
~;~~~
e^\phi |F_p|<1
~, 
}
in order  for the higher-order flux corrections to be under control.   In the second inequality above, the norm of the $p$-form flux is given by $|F_p|^2:=\frac{1}{p!}|F_{M_1\cdots M_p}F_{N_1\cdots N_p}g^{M_1N_1}\dots g^{M_pN_p}|$.

Conditions \eqref{cond1}, \eqref{cond2}, \eqref{cond3} are necessary and sufficient for the supergravity solution  given in \eqref{metricnh}, \eqref{flux}, \eqref{dilaton} to be 
within its regime of validity.

\subsection{Scale separation }

Let us first note that any rescaling of the $v_i$'s can be cancelled by a corresponding rescaling in $g_s$.  
So in the following we can keep  $v_i$ fixed without loss of generality. 

Suppose there is a solution parameterized by $\{g_s, n_a, v_i\}$.  
Let us moreover rescale, 
\eq{
n_0\rightarrow p~\! n_0~;~~~  n_{1,2,3}\rightarrow q ~\! n_{1,2,3}~;~~~  
g_s 
\rightarrow t\cdot q^{-1}~\! g_s
~, 
}
for some non-vanishing $p,q \in\mathbb{N}^*$, $t\in \mathbb{R}_+$.  
Under this rescaling we have, 
\eq{\spl{
 L_4  & \rightarrow   t ~\!\Big(\frac{p}{q}\Big)^{\frac14}   L_4
~;~~~ 
L_6\rightarrow \Big(\frac{p}{q}\Big)^{\frac14}    L_6~,~~~ 
\frac{L_4}{L_6}
\rightarrow
t ~\!\frac{L_4}{L_6}
~;~~~ 
T_i\rightarrow    \Big(\frac{p}{q}\Big)^{\frac12}  T_i \\
e^\phi    &\rightarrow    t ~\! q ^{-1}  \Big(\frac{p}{q}\Big)^{\frac34}    e^\phi 
~;~~~ 
e^\phi   |F_2| \rightarrow    t ^{-1}  \Big(\frac{p}{q}\Big)^{-\frac14}    e^\phi   |F_2|
~,~~~ 
e^\phi   |F_6| \rightarrow    t ^{-1} \Big(\frac{p}{q}\Big)^{-\frac14}    e^\phi   |F_6|
~,
}}
where $i=1,2,3$. 
Scale separation ($L_4\gg L_6$) is thus equivalent to taking $t\gg 1$. 

If in addition we want to respect conditions  \eqref{cond1}, \eqref{cond2}, we also must take $p\gg q$. Then the second of the two inequalities in \eqref{cond3} 
is automatically satisfied. To satisfy the first inequality in \eqref{cond3}, it suffices to take $p=q^2\gg 1$, and $t=q^{r}$, with $0<r<\frac14$. We are then guaranteed to be within the validity regime of the supergravity approximation. 
 

\subsection{Comparison with \cite{Bonnefoy:2019nzv}}

In order to compare with \cite{Bonnefoy:2019nzv} let us first  redefine the constant $g_s\rightarrow g_s (v_1v_2v_3)^{-\frac12}$, and also set $l_s=1$, so that  the formulae of 
section \ref{sec:nh} become,
\eq{\label{nf}
\frac{L_4}{L_6}=
 { 2\pi^2 ~\!g_s}
(n_1n_2n_3)^{\frac13} 
~;~~~
e^\phi=g_s(2\pi)^3\Big(
\frac{n_0^3}{n_1n_2n_3}
\Big)^{\frac{1}{4}} 
 ~,
}
and the 4d dilaton reads,
\eq{
e^{\phi_4}
= \frac{e^\phi}{\sqrt{V}}=g_s
~,}
where $V:=L_6^{6}$.

We see that the dependence on the 2-tori areas $v_i$ disappears, having been absorbed in the independent constant $g_s$. This is consistent with the attractor mechanism according to which the 
near-horizon geometry is fixed by the charges, and in particular is independent of the values of the K\"{a}hler moduli at asymptotic infinity. The latter correspond to the areas of the 2-tori, $v_i$. 
On the other hand,  the values of  the K\"{a}hler moduli at the horizon are given in \eqref{deft}, and correspond to the effective values of the areas of the 2-tori at the horizon. Indeed the $T_i$'s here are essentially the same as defined in eq.~(73) of \cite{Bonnefoy:2019nzv}.

The ratio $L_4/L_6$ here, cf.~\eqref{nf},   
corresponds to $m_{\text{KK}} (S_{\mathcal{N}=2})^{1/2}$ of \cite{Bonnefoy:2019nzv}, up to a numerical factor of order one. We find agreement with \cite{Bonnefoy:2019nzv}, 
cf.~eq.~(79) therein, provided we include the string coupling constant $g_s$ there.

\section{Generalizations}\label{sec:gen}

We will look for solutions of (massive) IIA supergravity the form $M_2\times M'_2\times M_6$, where $M_2$ is a two-dimensional maximally-symmetric Lorentzian manifold 
(i.e.~$\mathbb{R}^{1,1}$, dS$_2$, AdS$_2$), 
$M'_2$ is a two-dimensional maximally-symmetric Riemannian manifold (i.e.~$\mathbb{R}^{2}$, S$^2$, H$^2$) or discrete quotients thereof, and $M_6$ is a six-dimensional 
nearly-K\"{a}hler (NK), Calabi-Yau (CY), or Einstein-K\"{a}hler manifold. 
In \S\ref{sec:45} we will also consider the case where $M_6$ is a product of two-dimensional  Riemannian manifolds.

\subsection{$M_6$ nearly-K\"{a}hler with $m\neq0$}\label{sec:3.1}

The ansatz of the present section can be obtained from the consistent truncation of \cite{corfu} \S 5 therein,  by setting the scalars to constants and taking the one-forms therein to obey $\gamma=\chi\alpha$. We are following the conventions of that reference.
The ten-dimensional metric reads,
\eq{\label{tdma}\d s^2_{(10)} =  g_{\mu\nu}\d x^{\mu}\d x^{\nu}+ g_{ij}\d x^{i}\d x^{j} +g_{mn}\d y^m\d y^n 
~,
}
where $\{x^{\mu}$, $\mu=0,1\}$ are coordinates on $M_2$,  $\{x^{i}$, $i=2,3\}$ are coordinates on $M'_2$ and 
$\{y^{m}$, $m=1,\dots, 6\}$ are coordinates on $M_6$. The respective Ricci tensors are given by,
\eq{\label{rt}
R_{\mu\nu}=\Lambda_1g_{\mu\nu}~;~~~ R_{ij}=\Lambda_2g_{ij}~;~~~ R_{mn}=\Lambda g_{mn}
~,}
where the signs of $\Lambda_{1,2}$ are unconstrained at this point. 
The  NK manifold $M_6$ possesses a real two form $J$ and a complex three-form $\omega$ obeying,  
\eq{\spl{
\d J&=-6\omega\mathrm{Re}\Omega \\
\d\mathrm{Im}\Omega&= 4\omega J\wedge J ~.
\label{torsionclassesbnk}
}}
With these conventions, the constant $\omega\in\mathbb{R}$ above is related to $\Lambda$ in \eqref{rt} via,
\eq{\label{lw}\Lambda=20\omega^2~.}
We will assume that the dilaton is constant. 
Moreover, our ansatz for the forms reads,
\eq{\spl{\label{foranscy0}
F&=  \beta+m\chi J~;~~~ H= -6\omega \chi \text{Re}\Omega \\
G&=\varphi\text{vol}_2\swed \text{vol}'_2+\frac12  \xi J\swed J+\chi J\swed  \beta
~,}}
where $\chi$, $\varphi$, $\xi$ are real constants, and $\text{vol}_2$, $\text{vol}'_2$ are the volume forms of $M_2$, $M'_2$ respectively.\footnote{
Our conventions for the volume form in $D$ dimensions are: $\text{vol}_D=\frac{1}{D!}
\varepsilon_{M_1\dots M_D}\d x^{M_1}\dots \d x^{M_D}$. In the case of  Lorentzian signature we assume $\varepsilon_{0\dots D-1}=+1$.} 
The two-form $\beta$ is given by,
\eq{\label{twoform}
\beta=-\left(
f\text{vol}_2+f'\text{vol}'_2
\right)
~,}
where $f$, $f'$ are real constants. 
It can then be seen that \eqref{foranscy0} automatically obeys the Bianchi identities \eqref{bi}.

Plugging the above ansatz into the ten-dimensional equations of motion we obtain the following: the internal $(m,n)$-components of the Einstein 
equations  read,
\eq{\spl{\label{et1}
\Lambda&= 
\frac{1}{16}(1+5\chi^2)~\!m^2 
+\frac{1}{16}(1+\chi^2)  (f^2-f^{\prime2})
 +18  \omega^2\chi^2
+\frac{3}{16} \varphi^2
+ \frac{7}{16} \xi^2
~.}}
The  $(\mu,\nu)$-components read,
\eq{\spl{\label{et2}
\Lambda_1&=
-\frac12 (1+3\chi^2)~\! f^2
+\frac{1}{16}  \Big[
(1+9\chi^2)(f^2-f^{\prime2})+
(1-3\chi^2)~\! m^2-5\varphi^2
-288\omega^2\chi^2
-9\xi^2
\Big]
~.}}
The  $(i,j)$-components read,
\eq{\spl{\label{et22}
\Lambda_2&=
\Lambda_1+\frac12(1+3\chi^2)
(f^2+f^{\prime2})
~,}}
where we have taken \eqref{et2} into account. 
All the mixed $(\mu,m)$, $(i,m)$, $(\mu,i)$ components are automatically satisfied. 

The dilaton equation reads,
\eq{\spl{\label{et3}
0&=
3(1+\chi^2)(f^2-f^{\prime2})
-(5+9\chi^2)~\!m^2+288\omega^2\chi^2
+\varphi^2-3\xi^2
~.}}
The $F$-form equation of motion is automatically satisfied. 
The $H$-form equation reduces to the following three equations,
\eq{\spl{\label{hfeom}
0=\xi\varphi-48\omega^2\chi-2m\xi\chi-m^2\chi+2ff'\chi^2+(f^2-f^{\prime2})~\!\chi
~,}}
and, 
\eq{\spl{\label{seqh}
0&=  -3f\xi\chi -f\varphi+mf'+3m\chi^2f' \\
0&=  3f'\xi\chi +f'\varphi+mf+3m\chi^2f
~.}}
The $G$-form equation of motion reduces to,
\eq{\spl{
\label{gfeom1}
\omega(\xi-\chi\varphi)=0
~.}}
For $\omega$, $m\neq0$ this system of equations implies $\xi=\chi\varphi$ and $f=f'=0$, from which we see in particular that 
$\Lambda_1=\Lambda_2$, 
so that no scale separation is possible.  
There are three classes of solutions, as given in \cite{lmmt} \S 11.4 therein.\footnote{\label{foot1}The solutions 
of \cite{lmmt} are of the form AdS$_4\times M_6$. However, 
this gives rise to the exact same equations of motion as in the present case: a space of the form AdS$_2\times$H$^2$ 
subject to $\Lambda_1=\Lambda_2$. Moreover we may replace H$^2$ by a discrete quotient thereof, i.e.~a Riemann surface $\Sigma_g$ of 
genus $g>1$. For a given curvature, the minimum volume is attained for $g=2$, cf.~\eqref{cveq} below.} Explicitly we have:

{\it First class}: 

\eq{\Lambda_1=\Lambda_2=-\frac{3}{2}m^2~;~~~
\Lambda=m^2
~;~~~
\varphi^2={5}m^2
~;~~~
\chi=0
~.}

\vfill\break

{\it Second class}: 

\eq{\Lambda_1=\Lambda_2=-2m^2~;~~~
\Lambda=\frac{5}{3}m^2
~;~~~
\varphi^2=3m^2
~;~~~
\chi^2=\frac13
~.}

{\it Third class}: 

\eq{\label{3c}\Lambda_1=\Lambda_2=-\frac{48}{25}m^2~;~~~
\Lambda=\frac{4}{3}m^2
~;~~~
\varphi^2=\frac{27}{5}m^2
~;~~~
\chi^2=\frac{1}{15}
~.}
We only expect the third class, given in \eqref{3c} above, to be supersymmetric: it can be obtained from the solutions 
of \cite{bc}, which are special cases of \cite{lt1}, by replacing AdS$_4$ by an AdS$_2\times$H$^2$ (or AdS$_2\times\Sigma_g$) 
space subject to $\Lambda_1=\Lambda_2$, cf.~footnote \ref{foot1}. A similar substitution of AdS$_4$ by an AdS$_2\times$H$^2$ 
(or AdS$_2\times\Sigma_g$) space can also be performed for all known AdS$_4$ solutions. The converse is not true, however, as the  AdS$_2\times$H$^2$ space 
allows for more general fluxes, which would otherwise break the symmetries of AdS$_4$.
 {As already mentioned, scale separation is not possible in any of the three classes of solutions above, since all  curvatures are of the same order.}

\subsection{$M_6$ nearly-K\"{a}hler with $m=0$}

The ansatz for the forms reads,
\eq{\spl{\label{foranscy0c}
F&=  \beta ~;~~~ H=  -6\omega\chi \text{Re}\Omega \\
G&=\varphi  \text{vol}_2\swed \text{vol}'_2+\frac12  \xi J\swed J+\chi J\swed  \beta
~,}}
with $\chi$, $\varphi$, $\xi$, $\beta$ as in \S \ref{sec:3.1}, and satisfies the Bianchi identities \eqref{bi} for $m=0$. The equations of motion are obtained from 
\eqref{et1}-\eqref{gfeom1} by  setting $m=0$ therein. 

In this case the equations of motion can be solved to give a one-parameter solution of the form AdS$_2\times$S$^2$, without scale separation,
\eq{\spl{-\Lambda_1&=\Lambda_2=\frac{3}{2}(f^2+f^{\prime2})~;~~~
\Lambda=\frac{1}{3}(f^{\prime2}-f^2)\\
\varphi=\xi&=0
~;~~~
\chi=\pm\sqrt{\frac{5}{3}}
~;~~~
f=\pm\frac{1}{27}(4\sqrt{69}-5\sqrt{15})f'
~,}}
where the sign of $\chi$ is correlated with the sign of $f/f'$. 
Note that $|f^{\prime}/f|<1$ as it should, since $\Lambda>0$ for a nearly-K\"{a}hler manifold, cf.~\eqref{lw}.
 {Scale separation is again not possible.}

\subsection{$M_6$ K\"{a}hler Einstein}\label{sec:ke}

The manifold $M_6$ is now assumed to be K\"{a}hler-Einstein with K\"{a}hler form $J$,  $\d J=0$. The form ansatz reads,
\eq{\spl{\label{foranscy0c}
F&=  \beta+\chi J~;~~~ H= 0 \\
G&=\varphi  \text{vol}_2\swed \text{vol}'_2+\frac12  \xi J\swed J+ J\swed  \gamma
~,}}
with $\chi$, $\varphi$, $\xi$, $\beta$ as before, while the two-form $\gamma$ is given by,
\eq{\label{twoformc}
\gamma=-\left(
g\text{vol}_2+g'\text{vol}'_2
\right)
~,}
where $g$, $g'$ are real constants. 
It can then be seen that \eqref{foranscy0c} automatically obeys the Bianchi identities \eqref{bi}.

Plugging the above ansatz into the ten-dimensional equations of motion we obtain the following: the internal $(m,n)$-components of the Einstein 
equations  read,
\eq{\spl{\label{et1bc}
\Lambda&= 
\frac{1}{16}m^2+\frac{5}{16}\chi^2 
+\frac{1}{16}  (f^2-f^{\prime2})
+\frac{1}{16}  (g^2-g^{\prime2})
+\frac{3}{16} \varphi^2
+ \frac{7}{16} \xi^2
~.}}
The  $(\mu,\nu)$-components read,
\eq{\spl{\label{et2bc}
\Lambda_1&=
-\frac{7}{16}  f^2
-\frac32  g^2
+\frac{1}{16}  \Big[
-f^{\prime2}+
9(g^2-g^{\prime2})+
m^2-3\chi^2-5\varphi^2
-9\xi^2
\Big]
~.}}
The  $(i,j)$-components read,
\eq{\spl{\label{et22bc}
\Lambda_2&=
\Lambda_1+\frac12 
(f^2+f^{\prime2})
+\frac32 
(g^2+g^{\prime2})
~,}}
where we have taken \eqref{et2bc} into account. 
All the mixed $(\mu,m)$, $(i,m)$, $(\mu,i)$ components are automatically satisfied. 

The dilaton equation reads,
\eq{\spl{\label{et3bc}
0&=
3 (f^2-f^{\prime2})
+3 (g^2-g^{\prime2})
-5m^2-9\chi^2
+\varphi^2-3\xi^2
~.}}
Both the $F$-form and $G$-form equation of motion is automatically satisfied. 
The $H$-form equation reduces to the following three equations,
\eq{\spl{\label{hfeombc}
0=\xi\varphi -2\xi\chi-m\chi+2gg'+(fg-f'g') 
~,}}
and, 
\eq{\spl{\label{seqhbc}
0&=  mf +f'\varphi+3\chi g+3g'\xi\\
0&=  mf' -f\varphi+3\chi g' -3g\xi
~.}}
One way to solve this system of equations would be to 
view \eqref{hfeombc}, \eqref{seqhbc} as a linear system of 
three equations for three unknowns $f$, $f'$, $\chi$.\footnote{The cases for which the system \eqref{hfeombc}, \eqref{seqhbc} does not admit solutions for $f$, $f'$, $\chi$ are special and 
must be considered separately.} The solution can then be 
substituted into  \eqref{et3bc} to obtain one constraint 
on the remaining unknowns: $g$, $g'$, $m$, $\varphi$, $\xi$. 
Equations \eqref{et1bc}-\eqref{et22bc} then simply determine the curvatures 
$\Lambda_1$, $\Lambda_2$, $\Lambda$.

Let us now examine whether we can obtain a hierarchy between the curvature scales. 
From  \eqref{et1bc}-\eqref{et3bc} we obtain,
 \eq{\spl{\label{3.38}
 \Lambda&=-\frac13(\Lambda_1+\Lambda_2)
 \\
 &=\frac16(
 \varphi^2+m^2)+\frac12(\chi^2
+\xi^2)
 ~.}}
 Moreover from \eqref{et22bc}, \eqref{3.38} it follows that $\Lambda_1\pm \Lambda_2\leq0$, so that,  
 \eq{\label{3.39}
 \Lambda_1\leq0~;~~~|\Lambda_1|\geq|\Lambda_2|
 ~.}
 If $\Lambda_1$, $\Lambda_2\neq0$, this then implies that the 
 external space $M_2$ 
 is at 
least as highly curved as the  internal space $M'_2$.

$\bullet$ If $\Lambda_1=0$ then, as we can see immediately from from \eqref{et22bc}, \eqref{3.38}, also $\Lambda_2$, $\Lambda$ vanish as well as all flux, 
and the solution reduces to empty $\mathbb{R}^{1,3}\times M_6$ space, with $M_6$ Ricci-flat. 

$\bullet$ If $\Lambda_2=0$ then from  \eqref{3.38} it follows that $|\Lambda_1|=3\Lambda$. Hence the 
curvatures of the  external space $M_2$ and the  internal space $M_6$ are of the same order.

$\bullet$ The equations of motion can easily be solved (e.g.~numerically) for 
$|\Lambda_1|\approx 3\Lambda\gg |\Lambda_2|$, so that the spaces $M_2$, $M_6$ are much more highly 
curved than $M_2'$. 
Of course this is the ``wrong'' kind of scale separation.

\subsection{$M_6$ Ricci-flat}\label{sec:rf}

As a special case of \S \ref{sec:ke} we can impose that $M_6$ is Ricci-flat, $\Lambda=0$, so that 
\eqref{3.38}  implies,
\eq{\label{3.40}m=\varphi=\chi=\xi=0~.}
The remaining equations of motion can then be solved to give a two-parameter solution without scale separation,
\eq{
-\Lambda_1=\Lambda_2=\frac{1}{4}(f^2+f^{\prime2})+\frac34(g^2+g^{\prime2})
~,}
where the constants $f$, $f'$, $g$, $g'$ are constrained to obey,
\eq{\label{cons}
f^2-f^{\prime2}+g^2-g^{\prime2}=0~;~~~fg-f'g'+2gg'=0
~.}
The solution corresponding to the brane configuration of section \ref{sec:rn} is a special (supersymmetric) case of the above solution: 
as we can see by comparing \eqref{flux} with \eqref{foranscy0c}, taking \eqref{twoform},  \eqref{twoformc}, \eqref{3.40} into account, 
it corresponds to setting $f'=g=0$. Indeed this is a solution of \eqref{cons} above, provided $f=\pm g'$.

In this case, to make contact with \S \ref{sec:rn}, we can associate to the curvatures $\Lambda_1$, $\Lambda_2$, radii $L_2$, $L_2'$ via:  
$|\Lambda_1| =\frac{1}{L^{2}_2}$, $\Lambda_2=\frac{1}{L^{\prime2}_2}$, and   $L_4:=L_2=L_2'$.  
Moreover,  in the case where $M_6$ is Ricci-flat,  the radius  $L_6$ is a free parameter 
(only subject to flux quantization) and does not enter the equations of motion \eqref{3.40}-\eqref{cons}.

\subsection{$M_{10}=M^{(1)}_{2}\times\dots\times M^{(5)}_{2}$}\label{sec:45}

Let us now consider ten-dimensional spacetimes of the form $M^{(1)}_{2}\times\dots\times M^{(5)}_{2}$, where $M^{(1)}_{2}$ is a two-dimensional maximally-symmetric space 
of Lorentzian signature whereas $M^{(i)}_{2}$, for $i=2,\dots,5$, are two-dimensional  maximally-symmetric spaces 
of Euclidean signature, or discrete quotients thereof. 
The solutions we present here  generalize  those in \cite{o1,o2},  in which the choice of flux is not the most general.

Let us set,
\eq{
R_{\mu\nu} = \Lambda_1g_{\mu\nu}~;~~~R_{mn(i)}  = \Lambda_i g_{mn(i)} 
~,}
where we have denoted by $R_{mn(i)}$, $g_{mn(i)}$ the Ricci tensor,  resp.~the metric components along $M_2^{(i)}$, $i=2,\dots, 5$. 
The form ansatz will be taken to be,
\eq{\spl{\label{foransprod}
F&=  -\sum_{i=1}^5f_{(i)}\text{vol}^{(i)} ~;~~~ H= 0 \\
G&=\frac12\sum_{i,j=1}^5g_{(ij)}\text{vol}^{(i)}\swed\text{vol}^{(j)}
~,}}
where we have denoted by $\text{vol}^{(i)}$ the volume element of $M^{(i)}_{2}$, and 
$f_{(i)}$, $g_{(ij)}$ are constants obeying: $g_{(ij)}=g_{(ji)}$, $g_{(ii)}=0$. 
This gives,
\eq{\spl{
F^2_{mn(i)}&= f_{(i)}^2  g_{mn(i)}~;~~~ F^2_{\mu\nu} = -f_{(1)}^2 g_{\mu\nu} ~;~~~
F^2 =   2\Big(   - f_{(1)}^2 +\sum_{i\neq 1} f_{(i)}^2 \Big)    \\ 
G^2_{mn(i)} &= 6\Big(   - g_{(i1)}^2 +\sum_{j\neq 1} g_{(ij)}^2 \Big) g_{mn(i)}~;~~~ G^2_{\mu\nu} = -6  g_{\mu\nu}   \sum_{i} g_{(i1)}^2   \\
G^2 &=  24\Big(    -\sum_{i}g_{(1i)}^2 +\sum_{1\neq i<j} g_{(ij)}^2  \Big) 
~,}}
It is also useful to list the Hodge duals,
\eq{\spl{\label{hdd}
\star F&=f_{(1)}\widehat{\text{vol}}{}^{(1)}-\sum_{i\neq 1}f_{(i)}\widehat{\text{vol}}{}^{(i)}\\
\star G&=-\sum_i g_{(1i)}\widehat{\text{vol}}{}^{(1i)}+\sum_{1\neq i<j}g_{(ij)}\widehat{\text{vol}}{}^{(ij)}
~,}}
where we have denoted $\widehat{\text{vol}}{}^{(i)}:=\tfrac{\text{vol}_{10}}{\text{vol}^{(i)} }$, 
$\widehat{\text{vol}}{}^{(ij)}:= \tfrac{\text{vol}_{10}}{\text{vol}^{(i)}  \swed \text{vol}^{(j)}  }$, 
$\text{vol}_{10}:=\text{vol}^{(1)}  \swed\dots\swed \text{vol}^{(5)}$.

The equations of motion are as follows: the Einstein equations reduce to,
\eq{\spl{\label{eeqg}
\Lambda_1&=\frac{1}{16}m^2-\frac{7}{16} f_{(1)}^2-\frac{1}{16} 
\sum_{i\neq1}f_{(i)}^2-\frac12\sum_{i}g_{(1i)}^2
-\frac{3}{16}\big(
\sum_{1\neq i<j}g_{(ij)}^2- \sum_i g_{(1i)}^2
\big)\\
\Lambda_i&=
\Lambda_{(1)}+\frac12\big(
f_{(1)}^2+f_{(i)}^2+\sum_{j\neq 1}g_{(ij)}^2
+\sum_{j\neq i}g_{(1j)}^2
\big)~;~~~i=2,\dots,5
~.}}
The dilaton equation reads,
\eq{\label{eeqgb}
0=3  \Big(   - f_{(1)}^2 +\sum_{i\neq 1} f_{(i)}^2 \Big)
    -\sum_{i}g_{(1i)}^2 +\sum_{1\neq i<j} g_{(ij)}^2   
+5m^2
~.}
Equivalently, the modified dilaton equation reads,
\eq{\label{fdkjhdfhjk}
\sum_{i}\Lambda_i=0~.
}
The equations of motion for the RR forms $F$, $G$ are automatically satisfied. The $H$-form equation of motion reads,
\eq{\spl{\label{hjkhkjhs}
 0&=mf_{(1)}+\sum_{i\neq 1} g_{(1i)} f_{(i)} -\big(
 g_{(23)} g_{(45)}+ g_{(24)} g_{(35)}+ g_{(25)} g_{(34)}
 \big)   \\ 
0 &= mf_{(i)}+\sum_{p\neq 1} g_{(ip)} f_{(p)} 
-  g_{(1i)} f_{(1)} 
+\big(
 g_{(1j)} g_{(kl)}+ g_{(1k)} g_{(lj)}+ g_{(1l)} g_{(jk)}
 \big) ~;~~~i=2,\dots,5
 ~,}}
where in the  second equation above it is assumed that $j<k<l$ and $j,k,l\neq 1, i$.

This system of equations can be solved in a similar fashion as that of \S  \ref{sec:ke}:\footnote{Note that the equations of motion of \S  \ref{sec:ke} can be recovered from \eqref{eeqg}-\eqref{hjkhkjhs} by setting,
\eq{\spl{
\Lambda_3=\Lambda_4&=\Lambda_5=\Lambda~;~~~
f_{(1)}=-f~;~~~f_{(2)}=-f'~;~~~f_{(3)}=f_{(4)}=f_{(5)}=-\chi
~;~~~
g_{(12)}=\varphi
\\
g_{(13)}&=g_{(14)}=g_{(15)}=g~;
~~~g_{(23)}=g_{(24)}=g_{(25)}=g'~;
~~~g_{(34)}=g_{(35)}=g_{(45)}=\xi
~.\nn}}
} 
in general we can solve the linear system of 
five equations \eqref{hjkhkjhs}  for the five unknowns $f_{(i)}$. The solution can then be 
substituted into  \eqref{eeqgb} to obtain one constraint 
on the remaining unknowns $g_{(ij)}$,  $m$. Equations \eqref{eeqg} then simply determine the curvatures 
$\Lambda_{(i)}$, while \eqref{fdkjhdfhjk} is automatically satisfied.

It can easily be seen that 
the system of equations \eqref{eeqg}-\eqref{hjkhkjhs} admits solutions such that the curvatures $\Lambda_i$ are not necessarily equal. 
However, similarly to the case of \S  \ref{sec:ke}, it is impossible to achieve $|\Lambda_1| < |\Lambda_i|$, for $i=2,\dots,5$. 
This can be seen as follows: 
equations \eqref{eeqg} can be solved for $\Lambda_1$,  $\Lambda_i$,  in terms of the fluxes. Then using the dilaton equation \eqref{eeqgb} we find
that $-\Lambda_1$ and $-\Lambda_1\pm\Lambda_i$, $i=2,\dots,5$, can all be expressed as sums  of squares, 
so that,
\eq{
\label{l1ineq}
\Lambda_1\leq0~;~~~
|\Lambda_1|\geq |\Lambda_i|
~,}
for all $i=2,\dots,5$. Therefore, assuming the $\Lambda_i$ are not all vanishing, we conclude that the AdS$_2$ radius is bounded above by at least one of the 
radii of the internal factors. 
If we set $\Lambda_i=0$ for $i=2,\dots,5$,  so that the internal space is flat and the radii of the internal factors become free parameters, 
then \eqref{fdkjhdfhjk} would also imply $\Lambda_1=0$, so that all the fluxes vanish and we obtain 
a solution with flat ten-dimensional  spacetime.

\subsubsection{Flux quantization}\label{sec:fq}

For the supergravity solutions  to be promoted to full-fledged solutions of the quantum theory, flux quantization must be imposed. 
For simplicity, let us set the $2\pi l_s=1$ in the following. 
For $i,j=2,\dots,5$, flux quantization constrains the constants  $f_{(i)}$, $g_{(ij)}$ in \eqref{foransprod} to obey, 
\eq{\label{f1}
f_{(i)}=\frac{n_i}{V_i}~;~~~g_{(ij)}=\frac{n_{ij}}{V_iV_j}
~,}
where $n_i,n_{ij}\in\mathbb{Z}$ and $V_i:=\int_{M_2^{(i)}}\!\text{vol}^{(i)}$ is the volume of $M_2^{(i)}$. The constants 
$f_{(1)}$, $g_{(1i)}$ are constrained to obey,
\eq{\label{f2}
f_{(1)}=\frac{n_1}{V_2\dots V_5}~;~~~g_{(1i)}=\frac{n_{1i}  V_i}{V_2\dots V_5}
~,}
where $n_1,n_{1i}\in\mathbb{Z}$. Moreover, the Romans mass is constrained to be an integer, $m\in\mathbb{Z}$.

Let us also note that for the two-sphere S$^2$, or a discrete quotient of the hyperbolic space H$^2$, the volume is related to the scalar curvature 
by  the Gauss-Bonnet theorem. 
Indeed, a (compact) Riemann surface $\Sigma_g$ of genus $g>1$ can be obtained as a discrete 
quotient of 
the two-dimensional (non-compact) hyperbolic space H$^2$, 
$\Sigma_g=$H$^2/\Gamma$, where $\Gamma$ is a discrete subgroup of $SO(1,2)$.  
Let the Ricci tensor of $\Sigma_g$ be given by  $R_{mn}=\Lambda g_{mn}$. 
The Gauss-Bonnet theorem then implies, 
\eq{\label{cveq}
|\Lambda|=4\pi(g-1)V^{-1}
~,}
where $V=\int_{\Sigma_g}\!\!\!\text{vol}$ is the volume of $\Sigma_g$. For the two-sphere S$^2$ the corresponding relation reads, 
\eq{\label{sq}
\Lambda=4\pi V^{-1}
~.}
We have not been able to find a solution with  general flux to the system of equations of motion subject to \eqref{f1}, \eqref{f2}. 
However, as we now show, special solutions are  possible for  $m=0$, $g_{(ij)}=0$, i.e.~for 
vanishing Romans mass and four-form flux.  
In this case the dilaton equation, 
\eqref{eeqgb} reads,
\eq{\label{eeqgexdil}
f_{(1)}^2=\sum_{i\neq1}f_{(i)}^2
~.}
Taking this into account, 
the Einstein equations \eqref{eeqg} read,
\eq{ \label{eeqgex}
\Lambda_1=-\frac{1}{2} f_{(1)}^2 ~;~~~
\Lambda_i=
 \frac12 f_{(i)}^2 
~,}
for $i=2,\dots, 5$, while the remaining equations of motion are automatically satisfied.

We have already discussed solutions, obeying flux quantization, of the form AdS$_2\times $S$^2\times$T$^6$. Let us instead suppose that 
the curvatures of the internal manifolds are all strictly positive, $\Lambda_i>0$, for $i=2,\dots,5$. Flux quantization, \eqref{f1}, \eqref{f2}, taking \eqref{sq} into account, implies,
\eq{
f_{(1)}=\frac{n_1}{(4\pi)^4}\Lambda_2\dots \Lambda_5~;~~~f_{(i)}=\frac{n_i}{4\pi}\Lambda_i
~,}
for $i=2,\dots, 5$. 
Then \eqref{eeqgex} solves for the curvatures in terms of the quanta,
\eq{
\Lambda_1=- \frac{(8\pi)^8n_1^2}{2n_2^4\dots n_5^4}
~;~~~
\Lambda_i=\frac{2(4\pi)^2}{n_i^2}
~,}
while the dilaton equation imposes,
\eq{
(8\pi)^6n_1^2 = n_2^4\dots n_5^4\sum_{i=2}^5\frac{1}{n_i^2}
~.}
This equation clearly does not admit any solutions for integer $n_i$. Therefore solutions of the form AdS$_2\times $S$^2\dots\times$S$^2$, while  
admissible in supergravity, are excluded in the quantum theory.

Let us now consider the case where 
the curvatures of the internal manifolds are  strictly positive, $\Lambda_i>0$, for $i=2,\dots,4$, while $\Lambda_5=0$. 
Unlike the previous case, now the volume $V_5$ of $M^{(5)}_2\!\!=$T$^2$ is not related to its curvature, and thus does not enter the equations 
of motion. 
Flux quantization now implies,
\eq{
f_{(1)}=\frac{n_1}{(4\pi)^3}\frac{\Lambda_2\Lambda_3 \Lambda_4}{V_5}~;~~~f_{(i)}=\frac{n_i}{4\pi}\Lambda_i
~;~~~f_{(5)}=\frac{n_5}{V_5}
~,}
where $i=2,3,4$.  
Then \eqref{eeqgex} solves for the curvatures in terms of the quanta,
\eq{
\Lambda_1=- \frac{(8\pi)^6n_1^2}{2V_5^2n_2^4n_3^4 n_4^4}
~;~~~
\Lambda_i=\frac{2(4\pi)^2}{n_i^2}
~;~~~
\Lambda_5= n_5=0
~,}
for $i=2,3,4$, 
leaving $V_5$ a free parameter. 
The dilaton equation imposes,
\eq{
(8\pi)^4n_1^2 =  V_5^2n_2^4n_3^4 n_4^4\sum_{i=2}^4\frac{1}{n_i^2}
~.}
This equation simply determines $V_5$ in terms of the flux quanta, and always admits a solution. 
Therefore solutions of the form AdS$_2\times $S$^2\times$S$^2\times$S$^2\times$T$^2$, are possible in the quantum theory. 
Note also that by taking large enough quanta we can make sure we are in the regime of small curvature and large volume (in string units). 
Moreover the dilaton is a free parameter of the solution, and can be tuned to weak coupling so as to ensure we remain within the validity regime of 
the supergravity approximation.

Similarly one can show that 
 solutions of the form AdS$_2\times $S$^2\times$S$^2\times$T$^4$, obey flux quantization. In the latter case we may also replace 
 T$^4$ by a K3 surface.

\section{Discussion}\label{sec:discussion}

We have investigated superstring and supergravity backgrounds of the form 
${\cal M}^{(1)}\times \dots \times {\cal M}^{(n)}$ with special emphasis on the question of whether or not scale separation between the different factors is possible. 
We have seen that in all our solutions the scalar  curvature of AdS$_2$  (in absolute value) must be of the same order or larger  than the curvatures of 
all  the other factors. Moreover, the other factors cannot all be (Ricci-)flat: in the solutions presented here this would also force the curvature of AdS$_2$ 
and all the flux to vanish.  One might  therefore  invoke the relation between the radius, $L$, and the curvature: $\Lambda\simeq\tfrac{1}{L^2}$, to conclude that 
the radius of AdS$_2$ will be of the same order or smaller than the radius of at least one of the other factors.

Aside from the fact that the relation between scalar curvature and radius is more involved than what the previous paragraph suggests (many different definitions of the ``radius'' of a space are possible), there is a caveat to the argument of the previous paragraph: 
taking (possibly singular) discrete quotients of the internal spaces considered here, would leave invariant their local properties  such as their curvature, while 
  changing their global properties such as the radius.   
I.e.~curvature hierarchies  only concern the local properties of the spaces and do not 
immediately translate to corresponding hierarchies of radii. One way to directly address the question of the radius of the internal space is to study the spectrum 
of the (scalar) Laplacian on that space, whose first non-vanishing eigenvalue in particular can serve to read off the radius. Indeed this was the approach used in \cite{tss,rtt} to establish the 
absence of scale separation in the vacua of \cite{lt3}.

For the supergravity solutions of \S\ref{sec:gen} to be promoted to full-fledged superstring solutions, flux quantization must be imposed. As we saw in section \S\ref{sec:fq}, 
 this is indeed possible to 
carry out in special cases, notably when the internal space includes a T$^2$, K3 or CY factor. However  the general problem seems rather involved and we have been unable to find a solution obeying flux quantization in the case of the most general flux ansatz. It would be interesting to examine whether this can be addressed algorithmically with the help of a computer. 

A possible issue with the supergravity solutions of \S\ref{sec:gen}  is their potential instabilities, given the fact that we expect them to be 
non-supersymmetric in general. For example the solutions of \S\ref{sec:3.1}, for which a supersymmetry analysis has been performed, come in three distinct classes only one of which is supersymmetric. On general grounds we would expect the non-supersymmetric solutions to be unstable \cite{ooguri,ov}. It would also be interesting to establish 
the supersymmetry (or absence thereof) of the 
remaining solutions of \S\ref{sec:gen}.

\vskip1.0cm\noindent
{\bf\Large Acknowledgements}

\vskip0.5cm
\noindent 
We would like to thank S. L\"ust, E. Palti and T. Van Riet for
for useful discussions. 
The work of D.L. is supported  by the Origins Excellence Cluster.
D.L. also like to thank the KITP in Santa Barbara and the Ecole Normale Superieure for hospitality, where 
part of this work was performed.

\vskip1.5cm
\appendix

\section{IIA supergravity}\label{sec:the10d}

Setting the fermions to zero, the  IIA action  reads, 
\eq{\spl{\label{action3}S= \frac{1}{2\kappa_{10}^2}\int\d^{10}x\sqrt{{g}}\Big(
&-{R}+\frac12 (\partial\phi)^2+\frac{1}{2\cdot 2!}e^{3\phi/2}F^2\\
&+\frac{1}{2\cdot 3!}e^{-\phi}H^2+\frac{1}{2\cdot 4!}e^{\phi/2}G^2
+\frac{1}{2}m^2e^{5\phi/2}\Big) 
+S^\mathrm{CS}
~,
}}
and  $S^\mathrm{CS}$ is the Chern-Simons term. 
The  equations of motion (EOM) following from  the action (\ref{action3}) read:

Einstein EOM,
\eq{\spl{\label{beomf2}
{R}_{MN}&=\frac{1}{2}\partial_M\phi\partial_N\phi+\frac{1}{16}m^2e^{5\phi/2}{g}_{MN}
+\frac{1}{4}e^{3\phi/2}\Big(  2F^2_{MN} -\frac{1}{8} {g}_{MN}  F^2 \Big)\\
&+\frac{1}{12}e^{-\phi}\Big(  3H^2_{MN} -\frac{1}{4} {g}_{MN}  H^2 \Big)
+\frac{1}{48}e^{\phi/2}\Big(   4G^2_{MN} -\frac{3}{8} {g}_{MN}  G^2 \Big)~,}}
where we have set: $\Phi^2_{MN}:=\Phi_{MM_2\dots M_p}\Phi_N{}^{M_2\dots M_p}$, for any $p$-form $\Phi$. 

Dilaton EOM,
\eq{\spl{\label{beomf1}
0&=-{\nabla}^2\phi+\frac{3}{8}e^{3\phi/2}F^2-\frac{1}{12}e^{-\phi}H^2+\frac{1}{96}e^{\phi/2}G^2 +\frac{5}{4}m^2e^{5\phi/2}
~.
}}
Combining the trace of \eqref{beomf2} with \eqref{beomf1}, we obtain the modified dilaton equation, 
\eq{\label{beomf1md}
0=2R-{\nabla}^2\phi-(\partial\phi)^2-\frac{1}{6}e^{-\phi}H^2
~.}
Form EOM's,
\eq{\spl{\label{beomf3}
0&=\d {\star}\big( e^{3\phi/2}F  )
+e^{\phi/2} H\swed {\star}  G  \\
0 &= \d{\star} \big( e^{-\phi}H\big)
 +e^{\phi}F\swed {\star}  G 
  -\frac{1}{2}G\swed G
+ e^{3\phi/2}m {\star} F  
\\
0&=\d
{\star} 
\big(
e^{\phi/2}G\big)
-H\swed G
~.
}}
The forms obey in addition the  Bianchi identities,
\eq{\label{bi}
\d F= mH~;~~~\d H=0~;~~~\d G=H\wedge F
~.}

\providecommand{\href}[2]{#2}\begingroup\raggedright

\endgroup

\end{document}